# Ultrafast laser-induced subwavelength structures towards nanoscale: the significant role of plasmonic effects

Min Huang,[1,*] Ya Cheng,[2] Fuli Zhao,[1] and Zhizhan Xu[2]

[1] *State Key Laboratory of Optoelectronic Materials and Technologies, Sun Yat-sen University, Guangzhou 510275, China*

[2] *State Key Laboratory of High Field Laser Physics, Shanghai Institute of Optics and Fine Mechanics, Chinese Academy of Sciences, P.O. Box 800-211, Shanghai 201800, China*

*Corresponding author, email: syshm@163.com*

Nowadays, plasmonics aiming at manipulating light beyond the diffraction limit has aroused great interest because of the promise of nanoscale optical devices. Generally, the ability of breaking diffraction barrier is achieved via controlling surface plasmons (SPs) on elaborate man-made structures as products of human ingenuity. Here, nevertheless, we demonstrate that in short-pulse laser ablation ultrafast active plasmonic structures spontaneously generate in virtue of plasmonic effects rather than human will. First, the splitting of laser-induced subwavelength gratings that is experimentally evidenced on ZnO, Si, and GaAs, is confirmed to originate in the conversion of SP modes from the resonant to the nonresonant mode and further to the inphase or antiphase asymmetric mode. Further, as pulse number increases the universal scaling-down of laser-induced nanostructures in a crater is derived from the conversion of physical regimes of plasmonic interaction from the optical to the electrostatic regime, which may arouse quasistatic SPs with interacting scales far beyond the diffraction limit and result in ultrafast, non-thermal ablation for extraordinary electrostatic enhancement. Generally, the plasmonic mechanisms reveal the physical essence of the link between the deep-subwavelength space-scale and the ultra-short timescale for ultrafast laser-induced nanostructures: basically, "nanoscale" tends to eliminate electromagnetic retardation effects and bring an instant respond to the incident field, and arouse electrostatic interactions with giant local-field enhancement that should facilitate the ultrafast ablation driven by tremendous electrostatic forces. Thus, active plasmonic structures provided with the characteristics of simultaneous "nanoscale" and "ultrafast" is apt to form spontaneously in ultrafast laser ablation. In addition, the plasmonic mechanisms can act as a powerful "evolutionary force" acting in multipulse ablation and promoting the growth of resonant nanostructures—various typical plasmonic structures may be self-generated through multipulse evolution, which can be considered as a kind of natural plasmonics. In short, plasmonic effects play an important role in ultrafast laser-induced subwavelength structures towards nanoscale, which would provide new insights into the nature of ultrafast laser-induced damage at deep-subwavelength scale.

## I. INTRODUCTION

Recent years, driven by the increasing trend towards nanotechnology, laser machining exceeding the diffraction limit and approaching nanoscale becomes an intriguing and challenging issue [1-4]. In virtue of the advantage of ultrafast laser, subdiffraction-limit spatial resolution can be achieved via nonlinear effects with precise fluence control in the focus [2-4], for instance the multiphoton polymerization [2,3]. However, it should be noted that such a subdiffraction-limit processing capacity of ultrafast laser is not only the crystallization of human wisdom, but also an intrinsic characteristic in ultrafast laser-induced damage [5-17]: ultrafast laser can spontaneously induce deep-subwavelength gratings (DSGs, generally $\Lambda/\lambda<0.45$, $\Lambda$: grating period, $\lambda$: laser wavelength) in various solid materials. The phenomenon is extremely important in terms of both technological and theoretical viewpoints—it shows a straightforward way to break the diffraction limit for laser micro/nano structuring and clarifies the singular deep-subwavelength characteristic that suggests underlying mechanisms beyond the classical theory [18]. In particular, the feature sizes of formed structures can be reduced to an astonishing 10-nm scale [13], merely a few tenth of $\lambda$, implying new physics for laser-induced damage at deep-subwavelength scale. That is, DSGs open up a window showing the underlying mechanisms of laser-induced nanostructures. However, such gratings that demonstrate a variety of morphological characteristics [5-17] would bring forth a good deal of possible mechanisms and puzzle researchers to obtain a consistent one. Up to now, the origin of the universal phenomenon is still an open question.

Besides DSGs, lately ultrafast laser-induced near-subwavelength gratings (NSGs, generally 1>Λ/λ>0.45) have attracted renewed interest [19-25] of the researchers, because of the application prospects [20] and the non-classical subwavelength characteristics [19,21-25] suggesting new mechanisms related to the excitation of surface plasmons (SPs) [22-25] beyond the traditional picture. Based on the principles of plasmonics, a concise physical picture for the formation of NSGs has been proposed [22]: NSGs result from the initial direct SP-laser interference and the subsequent grating-assisted SP-laser coupling, i.e., NSGs can be ascribed to a phenomenon of plasmonics. The new viewpoint rooting in plasmonics is of great help for understanding the subwavelength nature and positive feedback of NSGs.

Although DSGs and NSGs frequently appear simultaneously in the ablation crater [5-17], they are always ascribed to different origins and treated separately. Nevertheless, in some cases, Λ of DSGs appears near half of that of NSGs [15-17], or even DSGs may directly come from the splitting of NSGs [17]. Notably, the explicit splitting demonstrates a direct way for DSG formation and hints a plasmonic nature in considering the origin of NSGs, which can serve as an important clue for investigating the origin of DSGs. Actually, referring to the active field of plasmonics [26-41], the various SP modes sustained by artificial metallic gratings [26,28-32,34] may present actual models for us to probe into the phenomenon.

More generally, the splitting is only an epitome of the universal phenomenon—the scaling-down of ultrafast laser-induced structures towards nanoscale as pulse number ($N$) increases. However, the fundamental phenomenon has not received sufficient attention and treated seriously in previous studies—actually the underlying mechanisms responsible for the scaling-down phenomenon are just the key for comprehending the essence of deep-subwavelength features appearing in ultrafast laser ablation. That is, this is an interesting phenomenon worthy of special attention. Considering the SP-related field enhancement in the nanoapertures of laser high-excited surface [13] and the capability of nano-manipulating light of plasmonics [26-41], particularly the remarkable plasmonic mechanisms dominated at deep-subwavelength scale [36-38], we propose that the quasistatic SP mode [36] that can induce extraordinary field enhancement [36,37] in deep-subwavelength apertures would provide new insights into ultrafast laser-induced damage well beyond the diffraction limit.

Further, in previous studies few efforts have been made to explore the following basic question: why would deep-subwavelength structures appear spontaneously in ultrafast laser ablation? That is, what is the physical reason for the emergence of the subdiffraction-limit structuring capability along with the shortening of the duration of laser pulses to the femtosecond timescale? And moreover, what is the link between "nanoscale" and "ultrafast"? Virtually, the pronounced plasmonic characteristics shown by the various ultrafast laser-induced nanostructures should convey an important message for us to resolve the puzzles: spontaneous plasmonic nanoarchitectures in ultrafast laser ablation might greatly promote the ultrafast damaging process of materials at "deep-subwavelength" scale with certain "instant" field enhancement mechanisms, which in turn act as a strong "evolutionary force" for the growth of such structures. Linking "nanoscale" and "ultrafast", the standpoint based on plasmonics would help in clarifying the above issues.

On the other hand, generally, towards the emerging plasmonics the ability of nano-manipulating light benefits from the rapid advance of nano-manufacturing that can bring elaborate man-made nanostructures with unique optical properties, which make use of the effects of SPs by virtue of human ingenuity. On the contrary, distinct from the plasmonic architectures developed artificially, sophisticated photonic crystal structures composed of non-metallic mediums have been presented in nature for millions of years and highly optimized by lengthy evolution in living things like butterflies and beetles [42-44]. Analogous to natural photonic crystals, the spontaneous laser-induced nanostructures provided with prominent plasmonic characteristics could lead us to a new field—natural plasmonics, which is related to the plasmonic phenomenon occurring spontaneously without the aid of artificial structures. Interestingly, other than the natural photonic crystals formed via many millennia generation-to-generation evolution, the natural plasmonic nanostructures formed by ultrafast laser ablation displays pulse-to-pulse evolution in a short timescale. The viewpoint of natural plasmonics can give new insights into the evolution of ultrafast laser-induced nanostructures.

Moreover, in the multi-pulse evolution of ultrafast laser ablation, concentrating on a certain pulse impinging the pre-formed nanostructures, we can ascribe the physical process to the field of ultrafast active plasmonics [41]: during the pulse impinging, owing to the ultrafast excitation of abundant free electrons, no matter for semiconductor or dielectric, the ablated regions will undergo a strong, rapid change toward the physical properties, turn to be metallic, and thus be able to sustain plasmonic modes [13,22-25], which in turn seriously affect the mechanisms of ultrafast laser-matter interaction as introduced above. In detail, the dielectric constant of the high-excited instantaneous state of the irradiated material can be described by the Drude mode and mainly determined by the free electron density ($n_e$), which is related to the excited level of the irradiated material and may be varied in a wide range for different regions of the ablated crater due to the variational irradiation fluence coming of the Gaussian field distribution of laser beams. Actually, as demonstrated in the previous studies [6,8,13,14,16,17,21,22], the distribution of different kinds of gratings formed in a crater is strongly associated with the distribution of irradiation fluence, i.e., the distribution of $n_e$. Therefore, considering the ultrafast active nature of short-pulse laser-induced structures, in particular the deep-subwavelength structures that are closely related to certain ultrafast mechanisms, it is clear that the variable $n_e$ is a key parameter for us to understand the related phenomena concerning various ultrafast laser-induced structures, such as the above-mentioned grating splitting and scaling-down phenomena.

In the paper, with the experimental results on ZnO, Si, and GaAs, we provide explicit evidences for the grating splitting phenomenon that acts as a direct route for the formation of DSGs and the grating scaling-down phenomenon that further force DSGs to shrink towards nanoscale. Further, the comprehensive numerical studies based on the viewpoint of plasmonics would enlighten us on putting forward a clear physical picture for the formation of ultrafast laser-induced structures coming into nanoscale and clarifying the origin of the grating splitting and scaling-down phenomena.

## II. EXPERIMENTAL AND COMPUTATIONAL DETAILS

### A. Experimental details

In the experimental study, the dielectric and semiconductor crystals of ZnO (0001), GaAs (100), and Si (100) with optical polishing (surface roughness < 5 Å) were used as the samples of short-pulse laser ablation experiments.

For the ablation experiments, we paid close attention to the grating structures fabricated by linearly-polarized lasers at normal incidence with different wavelengths, fluences, and pulse duations. In detail, the experimental setup for laser ablation on material surfaces performed in ambient air was similar to that in Ref. [12]. A 125-fs Ti:Sapphire laser system (Spectra Physics Hurricane) with 800-nm central wavelength and variable repetition rate from 1 Hz to 1 kHz was used in the femtosecond experiments. Specially, 1280-nm wavelength for Si was exported by optical parametric amplification (OPA). Then, the linearly-polarized Gaussian laser beam, of which intensity was adjusted by a variable neutral density filter accurately, was focused onto the sample surface by a convex lens of 300-mm focal length with the applied pulse number controlled by an electric shutter. The samples mounted on an electric XYZ-translation stage were observed real-time through a CCD camera equipped with a long-focus objective lens. After laser processing, the morphologies of ablation craters were examined by scanning electron microscopes (SEMs, JEOL JSM-6380 and Quanta 400F).

### B. Computational details

The numerical simulations of optical characteristics of grating surfaces on ZnO irradiated by intense fs laser pulses at normal incidence with $E$-field parallel to the grating wave vector (TM polarization) are carried out by the methods of rigorous coupled wave analysis (RCWA) [45] and finite-difference time-domain (FDTD) [46]. For the simulations, the dielectric constant $\varepsilon$ of ZnO surface irradiated by intense ultrafast laser pulses, which exhibits a metallic characteristic, is represented by the Drude model

$$\varepsilon(\omega) = \varepsilon_c - \frac{\omega_p^2}{\omega(\omega + i\Gamma)}, \quad (1)$$

where $\varepsilon_c$ is the dielectric constant of material in normal state, $\omega_p = (e^2 n_e / (m_{ef} \varepsilon_0))^{1/2}$ is the plasma frequency, and $\Gamma = 1/\tau$ is the electron collision frequency ($\tau$ is the electron collision time). Based on the approach proposed in Ref. [22], the following simulation parameters for the high excited ZnO are used: $\varepsilon_c = 3.85$, $e = 1.6 \times 10^{-19}$ C, $m_e = 0.91 \times 10^{-30}$ kg, $m_{ef} = 0.19 m_e$, $\varepsilon_0 = 8.85 \times 10^{-12}$ F/m, and $\tau = 4.31 \times 10^{-15}$ s.

Via the RCWA method, the optical characteristics of the metallic grating can be represented by the calculations of reflectance ($R$) maps of the surface as functions of different grating and light parameters, such as $n_e$, $\Lambda$, $W$, $\lambda$, and the incident angle of light. Thus, various reflectance maps, such as the $n_e$-$\Lambda$ map and the dispersion map, or other evolving maps, such as the $n_e$-$W$ map, aiming at different physical aspects, can be calculated in a conformable manner. In the calculations, the reflectance is summed up from the zeroth to the fifth diffraction orders, $n_e$ is set in an appropriate range so that the SP effects are prominent [34], and $\Lambda$ is set in the range covering the scales of various gratings appearing in the experiments. In detail, the RCWA simulations of the $n_e$-$\Lambda$ and the $n_e$-W maps are all implemented with 800-nm wavelength at normal incidence; the number of orders for the RCWA calculations is set to be 20, high enough to obtain accurate results.

In the $n_e$-$\Lambda$ maps, we pay attention to the SP bands via calculating the reflectances of the grating surfaces. Such reflectance maps can provide us information about the absorption characteristics of the entire surface rather than the details of field distribution. However, in ablation we may be more concerned about the fields localized in the grooves, in particular for the cases in electrostatic regime. Actually, the field enhancement factor in the grooves can be estimated by a direct transformation of the reflectance map based on an intuitive assumption: the absorption of incident light almost completely occurs in the groove region, which is proportion to the *E*-field and *W* of the groove. With the assumption, we calculate the quantity $\Sigma = (1-R)/W$ in the $n_e$-*W* map instead of *R* in the $n_e$-$\Lambda$ map. The new quantity $\Sigma$ closely related to the *E*-field in the groove region is able to offer direct information about the local *E*-field enhancement when SP modes are excited.

For the simulations of field distributions of specific points in the reflectance maps, the FDTD method is used with the source setup of a monochromatic plane wave of TM polarization at normal incidence and the same material parameters as those in RCWA simulations. In the two dimension simulation box, the periodic boundary condition is imposed in x-direction (the $K_g$ direction), and the PML absorbing condition is imposed in z-direction (the direction of incident wave vector $k$ ($k = 2\pi/\lambda$)); the grid size is set to be 2 nm, fine enough for the deep-subwavelength structures studied in the paper. In addition, for a given simulation the time steps are set to be sufficiently large for obtaining a stable result. Actually, in the study the simulation results of FDTD are always consistent with those of RCWA, indicating the accuracy of the numerical results.

## III. GRATING SPLITTING PHENOMENON

### A. Experimental results of the grating splitting phenomenon

In ultrafast laser ablation, it is general that NSGs and DSGs appear simultaneously in different regions of the ablated crater [5-17]. However, it is always hard to obtain a definite relationship between them, because in terms of the entire crater, the period $\Lambda$ for a certain kind of gratings may vary in a wide range, especially for DSGs. Nevertheless, if we keep our eyes on certain local conversion regions of different kinds of gratings, an explicit relationship between NSGs and DSGs may be obtained, for instance the relationship of spatial frequency doubling accompanied via a simple, clever way—the gating splitting, shown as follows.

#### 1. Morphological characteristics of the grating splitting

Figure 1 demonstrates the typical examples of the grating splitting phenomenon occurring on ZnO, GaAs, and Si. By looking into the morphologies of the conversion regions, we can verify that the conversion from the coarse to the fine gratings may occur in a conformable manner—the new grooves emerge in the middle of the ridges, bringing on the grating splitting. Besides the splitting of NSGs, with a large *N* the splitting of DSGs may also arise, as shown in Fig. 1(b) where the secondary $\Lambda$ approaches the scale of 100 nm (see also Fig. S1 of Supplemental Material [47] for the splitting example referring to a large-area grating). In short, all the results directly evidence the grating splitting phenomenon, which will be further elaborated in the following part.

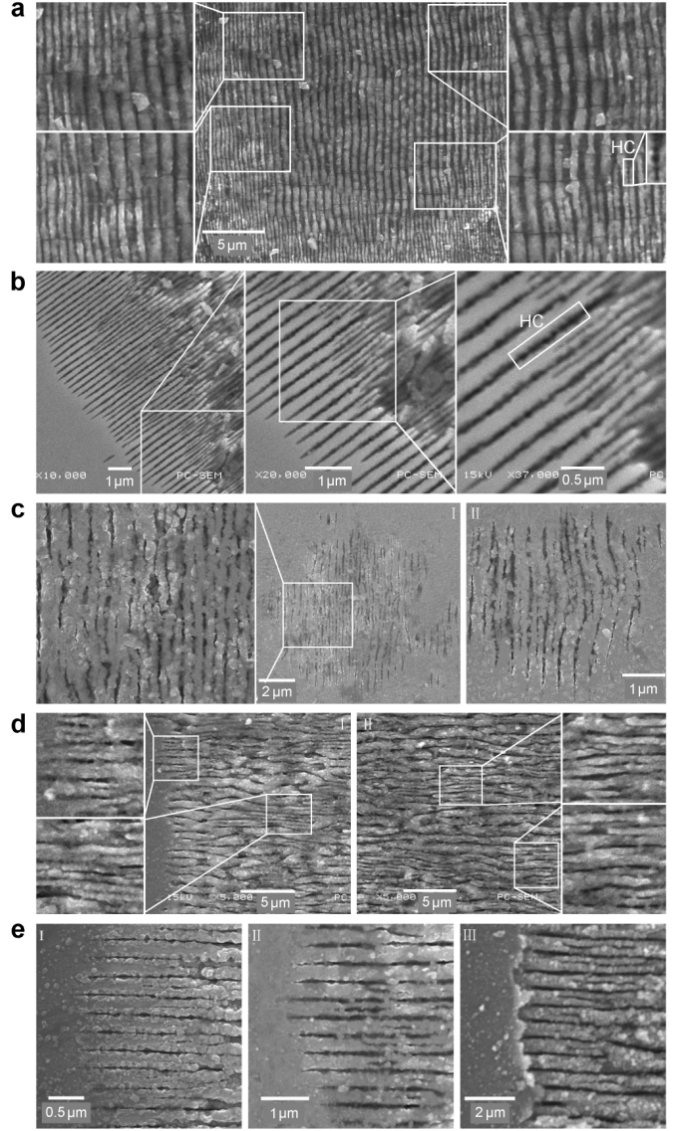

**FIG. 1. SEM images of the typical morphologies for the grating splitting phenomenon on different materials.** (a) ZnO surface irradiated by 800-nm fs laser with pulse number $N = 100$ at a fluence of 3.2 J/cm$^2$. Here the periods $\Lambda$ for NSGs and DSGs are 660 nm and 330 nm, respectively. (b) ZnO surface irradiated by 800-nm fs laser with pulse number $N = 5000$ at a fluence of 4.7 J/cm$^2$. Here the periods $\Lambda$ for coarse gratings and fine gratings are 280 nm and 140 nm, respectively, and the groove widths *W* for primary grooves and secondary grooves are about 35 nm and 15 nm, respectively. In (a) and (b), the areas denoted by "HC" show the hole-chain structures. (c) GaAs surface irradiated by 800-nm fs laser at a fluence of 0.1 J/cm$^2$ with a scanning technique [12]. The sub-figures I and II are both the locally-ablated areas in the irradiated surface. Here the period $\Lambda$ for the coarse gratings apt to split is 370 nm. (d) Si surface irradiated by 1280-nm fs laser with a scanning technique at the fluences of 0.16 J/cm$^2$ and 0.2 J/cm$^2$ in the sub-figures I and II, respectively. In these cases, the periods $\Lambda$ for NSGs in the respective splitting regions are 670 nm and 720 nm, obviously smaller than those in the non-splitting regions. (e) The cases of asymmetric splitting occurring at the peripheries of the ablated areas on ZnO, GaAs, and Si irradiated by fs laser with the wavelengths $\lambda$ of 800, 800, and 1280 nm, and the fluences of 2.8, 0.11, and 0.16 J/cm$^2$, respectively.

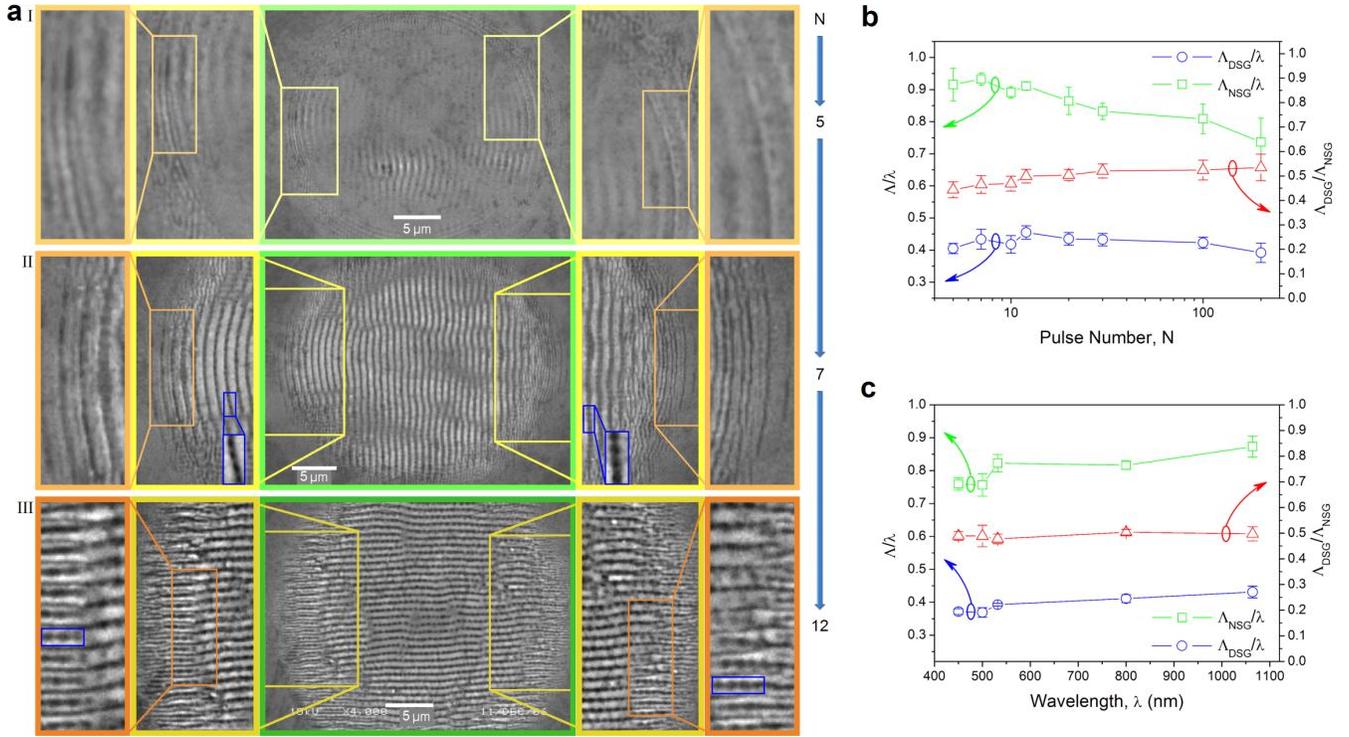

**FIG. 2. The evolution of the grating splitting phenomenon as the functions of pulse number $N$ and laser wavelength $\lambda$.** (a) SEM images of ZnO surface irradiated by 800-nm fs laser at a fluence of 2.6 J/cm$^2$ with different pulse numbers $N$ = 5, 7, and 12 corresponding to the cases I, II, III, respectively. In cases II and III, the areas surrounded by blue lines show the hole-chain structures. (b) Experimental dependence of pulse number $N$ vs grating periods of DSGs ($\Lambda_{DSG}$) and NSGs ($\Lambda_{NSG}$) and ratio $\Lambda_{DSG}/\Lambda_{NSG}$ in the conversion regions of craters on ZnO with the same irradiation condition as (a) (see (a) and Fig. S3 of Supplemental Material [47] for the corresponding SEM images). (c) Experimental dependence of laser wavelength $\lambda$ vs grating periods $\Lambda_{DSG}$ and $\Lambda_{NSG}$ and ratio $\Lambda_{DSG}/\Lambda_{NSG}$ in the conversion regions of craters on ZnO (see Fig. S4 of Supplemental Material [47] for the corresponding SEM images and irradiation conditions).

Now let's have a close look at the detailed splitting morphologies. Above all, there are obvious morphological characteristics for the grating provided with the tendency to split: generally, the grating exhibits wide, planar ridges and narrow grooves with sharp edges, which presents weak thermal features. In contrast, the grating being incapable of splitting always possesses a relatively-smooth profile, which displays strong thermal features. For instance, in Fig. 1(d) the appearance of the localized splitting gratings is different greatly from that of the non-splitting grating, which clearly indicates the close relationship between the grating morphologies and the splitting phenomenon. Such a relationship suggests that as $N$ increases, the evolution of the grating profile would lead to the grating splitting. Actually, the unique profile of the gratings bearing the splitting ability is quite surprising and inspiring—it forms spontaneously in ultrafast laser ablation and resembles that of the artificial metallic gratings fabricated for the interest of plasmonics, suggesting the plasmonic nature. In particular, the reduced thermal effects of the splitting regions deserve special attention: it means a lower excited level of the splitting regions accompanied with a smaller $n_e$ and a shorter heating time—an ultrafast, non-thermal process. That is, the split-ting tends to occur in the ultrafast, non-thermal ablation process, and in turn the fine gratings derived from splitting would facilitate the ultrafast, non-thermal ablation process. Conversely, the thermal effects coming of high irradiation fluence or certain intrinsic characteristics of materials, such as high absorption coefficient and thermal conductivity, may hinder the positive feedback of splitting. For example, due to higher thermal effects, in Fig. 1(d) for Si the splitting is harder to occur than in Fig. 1(c) for GaAs. Then, if we focus on the groove features of the splitting grating, we can see that the secondary grooves tend to be narrower than the primary ones, which leads to a compound grating made up of alternating wide and narrow grooves (Figs. 1 and 2(a)). It is worth noting that the newly-generated grooves may exhibit ultra-small widths ($W$) approaching the scale of 10 nm—the deep-subwavelength characteristic is highly interesting and inspiring, which implies the physical regime far beyond the diffraction limit and suggests a field enhancement mechanism dominating at nanoscale. With adequate feedback, the asymmetric grating will turn into being symmetric—the splitting process is accomplished. Moreover, further going deep into the groove morphologies of the gratings, we may dig out a substantial evidence of SP modes excited in the grooves: in many cases, apparently, hole-chain structures along the grooves can be easily observed (see Figs. 1(a), 1(b), and 2(a)), which seem to be the "footprints" of the plasmonic modes standing along the grooves in view of the field distributions of plasmonic waveguide modes [38-40].

### 2. Evolution of the grating splitting

Then, we would pay close attention to the evolution of the grating splitting. Generally, the splitting always initializes at the periphery of the ablated crater, as shown in Fig. 2(a). The result further confirms the dependence of the induction of grating splitting on $n_e$ that is monotonously related to the irradiation fluence—the damage-threshold fluence is propitious to the splitting. In detail, with a suitable fluence, the initial splitting occurring at the periphery merely requires a small pulse number $N$ (case I), and as $N$ increases, in the directions parallel (case II) and perpendicular (case III) to the NSG wave vector ($K_g$), the splitting both can develop. In addition, the splitting tends to be triggered by the increase of the groove depth ($D$) with multi-pulse feedback. As illustrated in Figs. 1(b) and 1(e) (see also Fig. S2 of Supplemental Material [47]), the splitting occurs slightly inside rather than at the outermost ablated region that is composed of a regular grating without splitting. This strange morphology is apt to occur with a large $N$, that is, with adequate feedback for the action of a large number of pulses. Actually, even with a small $N$, similar situations may also be observed: as the case II of Fig. 2(a) shown, after the action of a few pulses the explicit splitting is accompanied with the deepened grooves located slightly inside the ablated borderline instead of with the shallow, nascent grooves located at the outermost ablated region. These results demonstrate that the groove deepening with multi-pulse feedback is a prerequisite for the grating splitting. Further, as $N$ continues to increase, the splitting keeps on evolving along with the conversion regions moving towards the crater center, and eventually, all NSGs turn into DSGs provided $N$ is large enough. As shown in Fig. 2(b) (see Fig. 2(a) and Fig. S3 of Supplemental Material [47] for the corresponding SEM images), with $N$ increasing, concerning the conversion regions the ratio $\Lambda_{DSG}/\Lambda_{NSG}$ slightly rises around 0.5. Besides indicating an explicit spatial-frequency doubling relationship, the rising trend suggests the dominant gratings turning from NSGs to DSGs in the ablation process. In addition, with λ varying in the visible and near-infrared range, the splitting remains as a universal phenomenon. As shown in Fig. 2(c) (see Fig. S4 of Supplemental Material [47] for the corresponding SEM images), the ratio $\Lambda_{DSG}/\Lambda_{NSG}$ always locates near 0.5, explicitly revealing the spatial-frequency doubling relationship.

On the whole, at the initial stage of ultrafast laser ablation, NSGs are produced with higher priority and dominate in the crater, acting as the vanguard for the surface structuring. It is because that, at the stage NSGs can benefit the mechanism of SP-laser coupling, enhance the absorption of laser energy, and thus promote the ablation process [22]. Nevertheless, as above results exhibited, in ultrafast laser ablation with multipulse feedback, the formed NSGs are not always stable, especially for wide bandgap materials: as $N$ increases the conversion from NSGs to DSGs may occur due to the changes of $n_e$ and the grating profile (groove deepening and narrowing down). That is, the SP-laser coupling assisted by NSGs [22] is not invariably the dominant mechanism in ultrafast laser ablation—with the evolution of the ablation morphology, other electromagnetic modes favoring finer gratings with narrower grooves may prevail and further boost the ultrafast laser ablation via the nanostructuring process.

## B. Discussion of the grating splitting phenomenon

The above experimental results definitely show that the splitting mechanism can break through the diffraction limit and push laser-induced structures towards nanoscale. That is, the origin of the splitting phenomenon, which seems beyond the classical theory of laser-induced periodic surface structures, should be ascribed to certain fundamental electromagnetic mechanisms of laser-high-excited subwavelength gratings. Accordingly, in the following section, we have carried out a detailed numerical study on the issue.

### 1. Origin of the grating splitting phenomenon

In the detailed discussion, we start with the core issue: why may NSGs at the periphery split? In Ref. [22], the grating-assisted SP-laser coupling mode (recalled as the resonant $K_g$-SP mode here) has been proposed for the growth of NSGs. In order to resolve the issue about NSG splitting, naturally we should first dig into the feedback mechanism of NSGs related to the resonant $K_g$-SP mode. Above all, for the resonant $K_g$-SP mode, an important effect is that, as groove deepens the positive feedback mechanism forces Λ decreasing. In fact, such Λ-decreasing feedback requires a specific condition: the laser fluence is high enough, so that the thermodynamic feedback can keep up with the optical feedback. That is, for each pulse the thermal deformation or material ablation is strong enough to produce a sufficient change of the grating profile, thus the optical resonance condition can remains satisfied. Generally speaking, for a laser beam with a Gaussian field distribution, in the central region the fluence is high enough to maintain the feedback and promote the formation of NSGs; nevertheless, in the peripheral region with low fluence, the formed grooves are hard to be modulated for the melting layer being ultra-thin and the ablation efficiency being quite low, thus as the grooves further deepen, the thermodynamic feedback would lag behind the optical feedback. It means that in this situation NSGs are not in a state of positive feedback, that is, they are not stable. But so far, theoretically it is still unknown what kind of results such instability would lead to, which is just the key for us to understand the grating splitting phenomenon. For the reason, we complement a comprehensive electromagnetic analysis for the evolution of NSGs via the RCWA and FDTD simulations in order to answer the above-mentioned question.

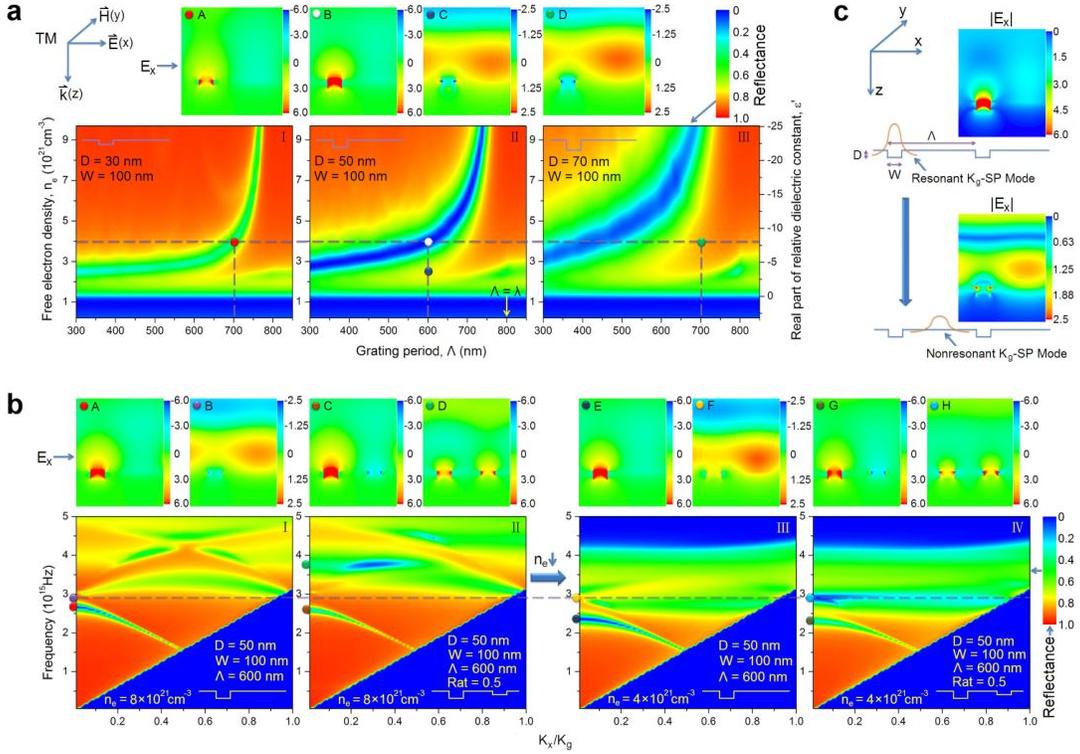

**FIG. 3. The simulations for the origin of the grating splitting phenomenon.** All the $n_e$-$\Lambda$ and dispersion maps simulated by RCWA method are towards the reflectances of the gratings with the parameters of grating profiles denoted in corresponding maps. The $E_x$-field distributions of certain points in the maps are simulated by FDTD method. (a) The $n_e$-$\Lambda$ maps of gratings with different groove depths $D$ aiming at laser wavelength $\lambda$ = 800 nm. Here, the $E_x$-field distributions of points A to D correspond to different SP modes in the $n_e$-$\Lambda$ maps. (b) The changes of dispersion maps due to the new groove emerging in the middle of the ridge for two different electron densities $n_e = 8\times10^{21}$ and $4\times10^{21}$ cm$^{-3}$. The parameter $Rat$ represents the ratio $D_n/D$ ($D_n$ is the depth of the new groove). Here, the $E_x$-field distributions of points A to H correspond to different SP modes in the dispersion maps at normal incident. (c) The physical picture for the initialization of grating splitting: the SP mode converts from resonant to nonresonant $K_g$-SP mode.

The $n_e$-$\Lambda$ maps concerning the reflectances of gratings with different groove depths $D$ on ZnO are shown in Fig. 3(a). In view of the above experimental facts, we propose that basically the splitting is caused by two direct reasons: groove deepening ($D$ increasing, from point A to point D) and dielectric constant $\varepsilon$ varying ($n_e$ decreasing, from point B to point C). Interestingly, the two paths regarding the loss of resonant $K_g$-SP mode have a similar result—a new electromagnetic mode appears with the high $E$-field located in the middle of the ridge. For the consistency with the experimental results, it is natural to conceive that this high $E$-field might produce localized damage, which should act as the seed for a new groove. However, before we arrive at a conclusion that in this manner a new grating with $K_g$ doubling ($2K_g$) has been produced, we must verify that the growth of new grooves is positive feedback. In other words, certain resonant electromagnetic modes are required to promote the growing of the new grooves. In the following section, referring to the fundamental electromagnetic interactions in metallic gratings, we concentrate on elucidating the physical origin of the grating splitting phenomenon.

Above all, in order to clarify the ridge-middle mode, we have carried out simulations for the dispersion maps of gratings with different situations (Fig. 3(b)). In plasmonics, the metallic grating is an important object of study [26,28-32,34], because it represents a basic diffraction model. Basically, as long as a grating profile contains higher harmonics in addition to the fundamental frequency, the $K_g$-SP interaction may give rise to a photonic band structure that exhibits a band-gap due to Bragg scattering [26,29]. In addition, for a grating with deep grooves, the waveguide resonances [28,29] (cavity modes) owing to localized SP and hybrid SP modes may greatly influence on the photonic band structure. In detail, for a grating provided with a dominant $K_g$ and an additional first harmonic $2K_g$, the $K_g$ component offers the momentum for the SP-light coupling whilst the $2K_g$ component opens up the band-gap, as shown in the dispersion maps of Fig. 3(b). Note that in the band-gap edge, the SP band is severely modified with the group velocity approaching zero—fields localize and extraordinarily enhance in the grooves (besides SP branches, the cavity mode presenting a hybrid character locate in the band-gap and fades away for normal incidence). In particular, the low energy SP branch is intensively excited (see points A and E in Fig. 3(b))—the origin of NSGs is just attributed to this branch [22]; nevertheless, the high-energy SP branch is weakly excited—it is not conducive to the formation of NSGs.

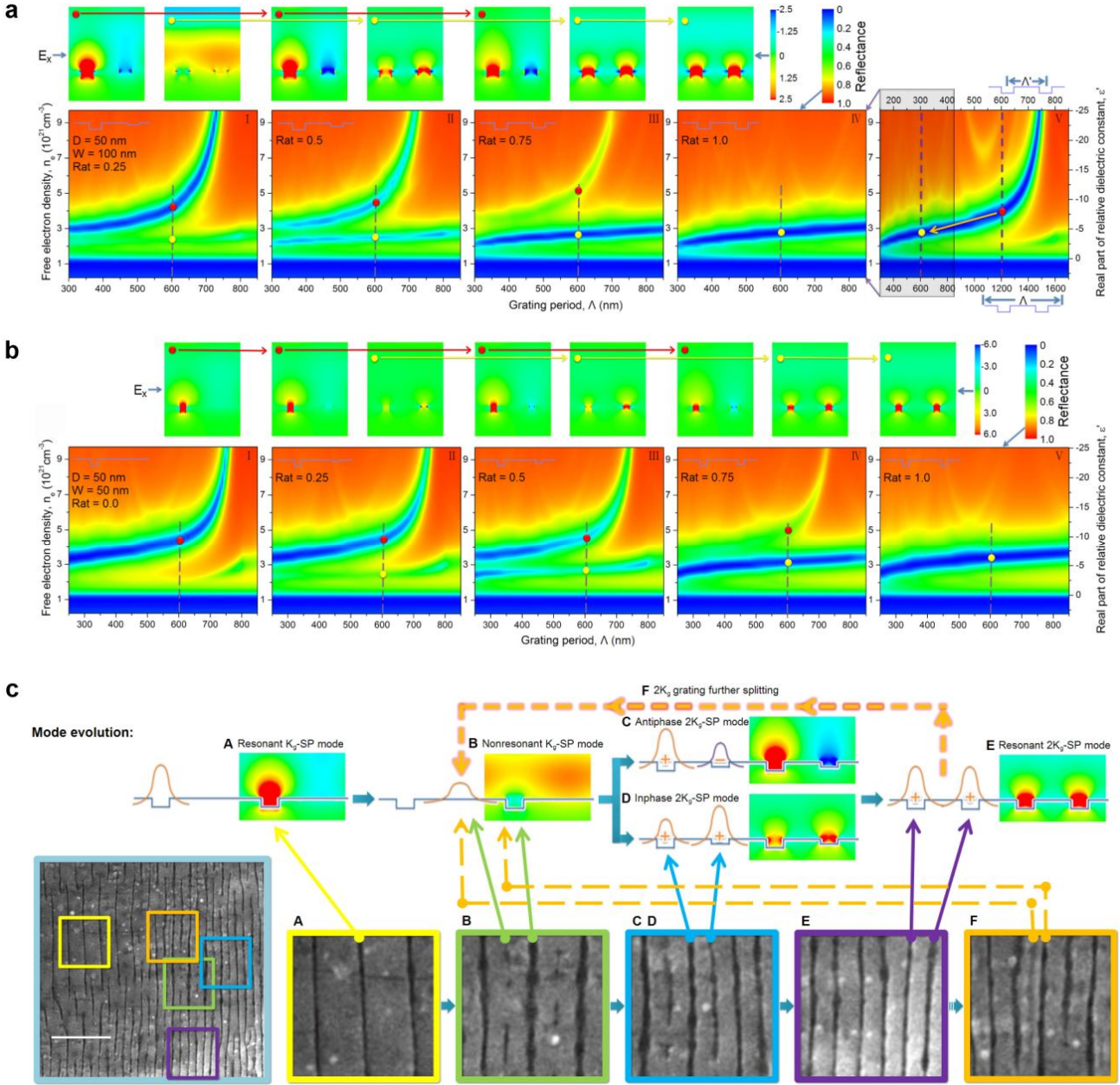

**FIG. 4. The simulations for the evolution of the grating splitting phenomenon.** All simulations aims at laser wavelength λ = 800 nm with grating-profile parameters denoted in the corresponding maps. The parameter *Rat* represents the ratio $D_n/D$. (a) As the new groove deepens from cases I to IV, the $n_e$-Λ maps demonstrate the evolution of SP modes. Specially, case V is the expansion of case IV in the grating-period range (the top tick label refers to the new grating). Here, $E_x$-field distributions of inphase (yellow points) and antiphase (red points) $2K_g$-SP modes with grating period Λ = 600 nm are provided for cases I to IV. Specially, in case V the red and the yellow points are corresponding to the original and the new gratings, respectively. (b) Resembling (a) but with a smaller groove width *W*. Note the difference of the color-map scales in (a) and (b). (c) Based on an experimental result on ZnO (irradiated by 800-nm fs laser with pulse number *N* = 500 at a fluence of 3.0 J/cm²) that exhibits an explicit splitting evolution, a complete physical picture for the grating splitting phenomenon is put forward. The scale bar is 2.5 μm.

Other than resonant $K_g$-SP mode (points A and E), the nonresonant region in the band-gap provide an important insight into the nonresonant situations in Fig. 3(a). As points B and F shown, in the band-gap the electromagnetic mode exhibits a high E-field located in the ridge middle, resembling the cases of points C and D in Fig. 3(a). Consequently, essentially, the ridge-middle mode in Fig. 3(a) can be considered as the band-gap mode. Actually, this electromagnetic mode has been proposed as the "nonresonant SP" mode in Ref. [31] (recalled as the nonresonant $K_g$-SP mode here). Thus, the SP-mode conversion from the resonant to the nonresonant $K_g$-SP mode gives rise to the initial splitting (Fig. 3(c)).

### 2. Positive feedback mechanisms for the grating splitting

As expected, the nonresonant $K_g$-SP mode will lead to a shallow ridge-middle groove. Then, the physical picture should evolve synchronously: the new grating provided with a more prominent $2K_g$ component is asymmetric, leading to certain interesting effects. Above all, the most prominent effect is the activation of a new SP branch (cases II and IV of Fig. 3(b)) near the original high-energy one, which presents the same phase between the E-fields of neighboring grooves (points D and H). Accordingly, we denote this mode as the inphase $2K_g$-SP mode, which inclines to promote the new groove deepening, especially for the low-$n_e$ case (point H). Meanwhile, along with the original low-energy SP branch slightly shifting down and weakening, the resonant $K_g$-SP mode turns into a new mode with an antiphase relationship between the E-fields of neighboring grooves (points C and G). Analogously, the mode is called the antiphase $2K_g$-SP mode. Thus, the further SP-mode conversion, such as from the nonresonant $K_g$-SP to the inphase $2K_g$-SP mode (from point B to point D, or point F to point H), may lead to positive feedback for the splitting. Actually, such a conversion tends to occur with a lower $n_e$ towards the plasma critical density ($n_c$) (from point F to point H, or point B to point H with $n_e$ decreasing), in order that the frequency conservation and stronger absorption can be achieved. Furthermore, the conversion from the resonant $K_g$-SP to the antiphase $2K_g$-SP mode coupled with $n_e$ slightly increasing seems also possible. In short, the dispersion maps illustrate the physical nature of positive feedback of the grating splitting: the evolution of SP modes as $N$ increases.

Now let's come back to the $n_e$-$\Lambda$ maps for the splitting evolution. In Fig. 4(a), resembling the dispersion maps, after a shallow ridge-middle groove emerges, besides the antiphase $2K_g$-SP mode (red points) corresponding to the original resonant $K_g$-SP mode, the new inphase $2K_g$-SP mode (yellow points) appears with a lower $n_e$ approaching $n_c$. In the actual ablation process, which mode is chosen should be determined by the instantaneous $n_e$; moreover, the conversion between the two modes is also possible. Then, with the new groove deepening (from case I to case IV), the inphase mode enhances and the antiphase one weakens. When the new groove grows to the depth of the original one (case IV), the antiphase mode eventually fades away, leaving the pronounced inphase one. It is worthy to note that virtually here the inphase $2K_g$-SP mode degenerates to a kind of resonant $2K_g$-SP mode, which is derived from the resonant $K_g$-SP mode with the $K_g$ doubling. As shown in case V with a larger $\Lambda$ range, a complete splitting from the red to the yellow point can be regarded as the conversion from the resonant $K_g$-SP to the resonant $2K_g$-SP mode via a "$\Lambda$-halving" (a discrete, "jumping" movement) in the resonant $K_g$-SP band. In contrast, the $\Lambda$-decreasing mechanism proposed in Ref. [22] can be considered as a continuous movement in the resonant band. This comparison demonstrates the wisdom of nature: the grating splitting mechanism provides a smart way for the further positive feedback of the ablation process while the direct $\Lambda$-decreasing mechanism is no longer in force.

Further, for a smaller $W$ in Fig. 4(b), although mainly the SP-mode conversion trend is similar to that in Fig. 4(a), here the dominant ranges of bands shifts towards smaller $\Lambda$ and $n_e$. Moreover, the E-fields are much more enhanced in the narrower grooves with a more prominent neighboring asymmetry, which may lead to more positive feedback for the splitting. Agreed with the experiments, these simulations reveal the fact that a finer grating matching with narrower grooves is prone to occur in the low-fluence region, where stronger field enhancement is required to achieve damage.

### 3. Physical picture for the grating splitting phenomenon

Then, referring to an experimental result that exhibits an explicit splitting evolution for the laser-induced gratings, we bring forward the complete physical picture for the grating splitting phenomenon. As shown in Fig. 4(c), the splitting evolution one-to-one corresponds to the SP-mode evolution. At first (case A), the resonant $K_g$-SP mode gives birth to NSGs. Then, along with the loss of the resonant $K_g$-SP mode (case B), the nonresonant $K_g$-SP mode takes effect and initializes the splitting—shallow, discontinuous grooves emerge in the middle of the ridges. In succession, the antiphase (case C) and inphase (case D) $2K_g$-SP modes play a vital role for the growth of the nascent grooves as $N$ increases. As long as the new grooves reach the depth of the original grooves (case E), the SP mode turns into the resonant $2K_g$-SP mode. Then as a loop (case F), a further splitting may arise following the same route from case B to case E. Note that commonly the secondary splitting bears a small possibility, because thermal effects tend to hinder the formation of ultrafine gratings. But at the periphery, where thermal effects are dramatically weakened (see Fig. 1(b)), with adequate feedback DSGs can further split into the 100-nm scale. In a word, the conversion of SP modes in the ablation is responsible for the grating splitting.

FIG. 5. **The scaling-down phenomenon of ultrafast laser-induced structures.** (a) For a typical crater on ZnO (irradiated by 800-nm fs laser with pulse number $N = 100$ at a fluence of 3.2 J/cm$^2$), the evolution details of various gratings from center to periphery is exhibited in the SEM images with abundant physical elements presented sequentially as denoted by A to H. (b) The $n_e$-$\Lambda$ maps concerning the effects of groove narrowing down without (cases I, II, and III) and with grating splitting (cases IV, V, and VI). Here $|E_x|$ distributions of points A to I correspond to different SP modes in $n_e$-$\Lambda$ maps with a grating period $\Lambda = 250$ nm. (c) The physical picture for the conversion of the physical regime of SP modes from the optical to the electrostatic regime in the cases of a flat metal-vacuum interface and a metallic slit. See Section II for the definition of the parameter X.

# IV. GRATING SCALING-DOWN PHENOMENON

## A. Experimental results of the grating scaling-down phenomenon

In the preceding part, we focus on the splitting phenomenon, a representative for the plasmonic effects acting on the ultrafast laser ablation. In fact, such a splitting is only an epitome of the more universal phenomenon: the scaling-down of ultrafast laser-induced structures as $N$ increases. In Fig. 5(a), for a typical crater, the evolution of various gratings from the central to the peripheral regions is exhibited with the related physical elements presented sequentially as follows. Above all, the central region is covered by a 630-nm-period NSG with 220-nm-width grooves and thick-amorphized ridges. Then outwards, a conversion region appears: NSGs split into DSGs. Here, coupled with the splitting, the asymmetry of the nascent fine grating and the narrowing-down of the newborn grooves (110 nm vs 30 nm for the widths of the neighboring grooves) can be clearly picked up. Further outwards, it is the peripheral region made up of pure DSGs. Obviously, towards the outmost region, DSGs with $\Lambda$ changing from 330 nm to 160 nm demonstrate a explicit decreasing trend, accompanied with the thin-amorphized ridges and the ultra-narrow grooves approaching 10-nm scale. Then at the outmost region, interestingly, certain fundamental grating components spontaneously come into being locally. First, the single groove of about 20-nm width appears in a self-sustaining manner, which is high significant considering the fact that via the conventional laser direct writing technique, it is almost impossible to produce an alone groove on such a deep-subwavelength scale. It seems that the SP modes propagating along the groove may contribute to the growth (deepening and extension) of the groove. Second, as the simplest DSG element, the asymmetrical free-standing pair of grooves with 200-nm spacing exhibits the nature of localized SP modes resembling the antiphase $2K_g$-SP mode. Third, as the simplest case of the grating splitting, the splitting at the middle of the pair emerges due to the loss of localized SP modes for the groove deepening. In short, these representative characteristics demonstrating abundant nanoplasmonic features would provide us significant insights for the origin of the surprising laser-induced structures at the deep-subwavelength scale.

As shown above, generally speaking, in multi-pulse ablation as $N$ increases the scaling-down of ultrafast laser-induced structures is a universal trend. For instances, the one-way conversion from the coarse to the fine gratings displays the process of $\Lambda$-reduction is positive feedback; the extraordinary narrowing-down of grooves, from the 100-nm scale in the high-fluence region to the 10-nm scale in the threshold-fluence region, is also a prominent evidence. Particularly, the ultra-narrow grooves that are conducive to form an ultra-regular grating and bring on the splitting strongly imply certain field enhancement mechanisms taking effect and being dominated on the deep-subwavelength scale, which would profit the ultrafast, non-thermal ablation. Hereinafter, we will conduct in-depth theoretical study to clarify this fundamental issue.

## B. Discussion of the grating scaling-down phenomenon

### 1. Effects of the groove width on the SP band

First, the numerical studies on the effects of groove scaling-down are carried out. From Figs. 4(a) and 4(b), it can be recognized that $W$ influences on the dominant region (high-absorption region) of the SP band and the enhancement of $E$-fields. The further comparing simulations in Fig. 5(b) can tell more details about these effects. As shown in cases I, II, and III, along with the groove narrowing down, the dominant region of the SP band in the $n_e$-$\Lambda$ maps shifts towards the deep-subwavelength region along with the obvious $E$-field enhancement in the grooves. Furthermore, the ratio $|E_x/H_y|$ also displays an explicit increasing, which means the enhancement of electrostatic effects. The simulations definitely suggest that the $W$ narrowing down coupled with the $\Lambda$ scaling down is propitious to the field enhancement in the grooves, and thus promote the grating growth in the low-fluence region. In addition, somewhat like the case of the groove deepening in Fig. 3(a), as the grooves narrow down the upshift of the SP band may lead to the loss of resonant modes and thus boost the grating splitting. Moreover, in cases IV, V, and VI, the simulations for the effects of the groove scaling-down accompanied with the grating splitting are performed. Compared with case II, in case IV since the new narrow groove arises, the dominant SP bands obviously shift towards a smaller grating scale. Moreover, with respect to Figs. 4(a) and 4(b), here the resonant $E$-field provided with a stronger asymmetry for the adjacent grooves is more localized and enhanced. Then, as the two adjacent grooves further scale down in company (case V), larger band widths, higher band intensities, and stronger $E$-fields appear with the modes further inclined to a smaller $\Lambda$. Here, the larger band width means stronger endurance to the coupling mismatch of SP modes caused by $n_e$ fluctuation, which should benefit the grating growth. However, in contrast to case V, case VI for the two grooves with an equal $W$ shows a weaker inphase mode with smaller $E$-field enhancement as well as a larger band gap that should increase the difficulty of the conversion between the inphase and the antiphase mode. That is, the narrowing-down of the new groove is advantageous for the grating splitting, which well agrees with the experiments.

### 2. Physical Regimes of SP Modes

Then we will probe into the physical nature for the groove narrowing-down phenomenon. Before we come into in-depth simulations, it would be of great help for us to learn more about plasmonics at the deep-subwavelength scale, at which the characteristics of SP modes change greatly and become abnormal: when $\lambda \gg W$, a kind of quasistatic SP mode [36] may be excited with the physical regime converting from the "optical regime" to the "electrostatic regime", which may bring on extraordinary $E$-field enhancement of several orders of magnitude and abnormal optical absorption.

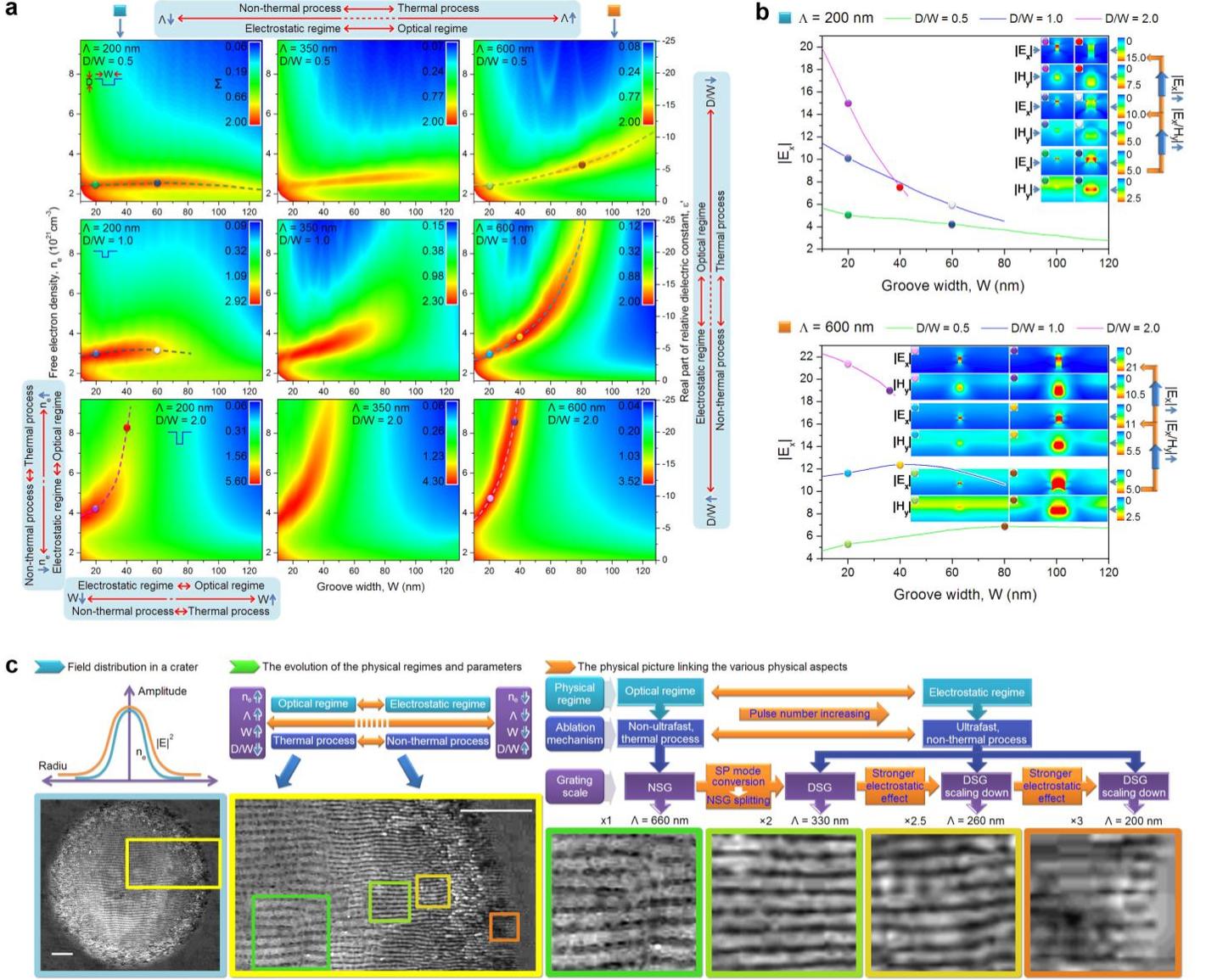

**FIG. 6. The origin of the scaling-down phenomenon and the physical picture for ultrafast laser-induced subwavelength structures towards nanoscale.** (a) The $n_e$-$W$ maps concerning the quantity $\Sigma$ (See Section II for the definition of $\Sigma$) of gratings with different values of $D/W$ and $\Lambda$ manifest the origin of the scaling-down phenomenon: the conversion of the physical regime from the optical to the electrostatic regime via a variety of routes, such as the decreases of $n_e$, $\Lambda$, and $W$, and the increase of $D/W$. Note the difference of the color-map scales in these cases. (b) The numerical results of the extraordinary $|E_x|$ enhancement in the middle of the groove open with the points and curves corresponding to the points and curves in (a). Here, the $|E_x|$ and $|H_y|$ distributions of the points of different colors correspond to the SP modes in the $n_e$-$W$ maps with different values of $D/W$ and $\Lambda$. Note the changes of color-map scales of different $D/W$ values. (c) Based on a typical crater on ZnO (irradiated by 800-nm fs laser with pulse number $N = 100$ at a fluence of 2.6 J/cm$^2$), the physical picture for the phenomenon of ultrafast laser-induced subwavelength structures towards nanoscale is presented. The scale bar is 5 μm.

In detail, the definitions of the physical regimes of SP modes are illustrated as follows. For the SP dispersion of a flat metal-vaccum interface (neglect the imaginary part of $\varepsilon$), there are two asymptotic regimes (see Fig. 5(c)): the "optical regime" with $\varepsilon$ approaching -∞, in which retardation effects dominate and the SP is provided with a behavior of light; the "electrostatic regime" with ε approaching -1, in which retardation effects can be ignored and the SP exhibits an electrostatic characteristic. For the normalized fields **E** and **H**, in the optical regime $|E_x/H_y|$ approximates 1, and in the opposite electrostatic regime $|E_x/H_y|$ approximates ∞ that corresponds to an extraordinary $E$-field enhancement. Referring to Ref. [36], a physical quantity $X = \delta_p / \delta_s$ was introduced to describe the relevant physical regimes, where $\delta_p$ is the penetration depth for SPs in the metal, and $\delta_s$ is the usual skin depth for a plane wave in the metal. In particular, for the flat interface, $\delta_p = 1/|k_s^\perp|$ ( $|k_s^\perp| = k|\varepsilon/\sqrt{\varepsilon+1}|$ is the modulus of the SP wave vector perpendicular to the interface), $\delta_s = 1/(\sqrt{|\varepsilon|}k)$, thus

$$X = \sqrt{(\varepsilon+1)/\varepsilon}. \quad (2)$$

With such a definition, the dimensionless parameter $X$, which satisfies $0 < X < 1$, can represent the two physical regimes in two extremes: $X \to 1$ for the optical regime and $X \to 0$ for the electrostatic regime.

Furthermore, for a metallic slit provided with $W < \lambda$, the same quantity $X$ can be introduced with an expression as

$$X = \sqrt{\frac{\varepsilon}{\varepsilon-1}} f(\Gamma), \quad (3)$$

Where $f(\Gamma) = -\Gamma + \sqrt{\Gamma^2 + 1}$ and $\Gamma = 1/(k|\varepsilon|\sqrt{|\varepsilon-1|}W)$. Then, interestingly, by reducing $W$ from the 100-nm to the 10-nm scale, the physical characteristic of SP modes in the groove can be turned from the optical regime to the electrostatic regime, as depicted in Fig. 5(c). Moreover, for the curves derived from different ε with metallic characteristics (ε < -1), ε approaching -1 ($n_e$ decreases towards $n_c$) is beneficial to the appearance of the electrostatic regime.

Actually, our above simulations in Fig. 5(b) just reflect these physical qualities: as the grooves narrow down, accompanied by the strong $E$-field enhancement the dominant SP band shifts to the smaller-Λ and lower-$n_e$ regime that exhibits a more prominent electrostatic characteristic. Accordingly, such a SP mode could enlighten us on the origin of the scaling-down phenomenon.

### 3. Origin of the grating scaling-down phenomenon

In the $n_e$-$W$ maps of Fig. 6(a), the quantity Σ (see the definition in Section II) related to the local $E$-field enhancement in the groove is calculated. With regard to the maps in the same row, it is obvious that as $D/W$ increases, the dominant SP band shifts towards a smaller $W$ accompanied by strong $E$-field enhancement (see the numerial results of Fig. 6(b) for the $E$-field amplitude at the middle of the groove open). Consistent with the experiments, the simulations indicate that the groove deepening is always coupled with the groove narrowing down. Moreover, under the same groove profile ($D/W$ keeps constant), as Λ decreases a similar trend can be observed: the dominant SP band shrinks towards a smaller $W$. The numerical results further demonstrate that the finer grating prefers a smaller $W$, which also agrees with the experimental results. On the whole, the simulations exhibit a variety of routes for the conversion of the physical regime from the optical to the electrostatic regime: $W$ decreases towards the 10-nm scale; $n_e$ decreases approaching $n_c$; Λ decreases towards the 100-nm scale; $D/W$ increases.

Furthermore, considering the extraordinary $E$-field enhancement from the optical to the electrostatic regime as well as the experimental fact of the weak thermal effects of the DSG regions, we propose that the electrostatic regime is accompanied with an ultrafast, non-thermal ablation process [13], and the optical regime is accompanied with a non-ultrafast, thermal ablation process. Such relevance between the physical regimes and the ablation mechanisms is straightforward: in the electrostatic regime, retardation effects of the electromagnetic field can be considered negligible, that is, the interaction of the matter with the electromagnetic field is reduced to a purely electrostatic one in an "instantaneous" response manner described by the electrostatic limit of Maxwell's equations [48]. In other words, the elimination of the electromagnetic retardation should facilitate the ultrafast ablation process. In details, once the ultrashort laser pulse impinges on a nanostructured surface, the nanostructures will respond to the incident field "instantaneously"—instant strong electric fields would arise in the nanoapertures and bring on ultrafast ablation driven by tremendous electrostatic forces, such as the ablation mechanism of Coulomb explosion [13], which exhibits extremely small thermal effects. By contrast, in the optical regime, the pronounced retardation effects would give rise to longer kinetics for the excitation of well resonant SP modes, that is, a longer duration of action (heating time) on material surfaces. In addition, the higher $n_e$ also means a higher electron temperature and longer decay dynamics of the excited electrons, both corresponding to stronger thermal effects. Thus, the optical regime conduces to the non-ultrafast, thermal ablation process.

### 4. Physical picture of the grating scaling-down phenomenon

At last, with a typical ablation crater on ZnO, we elucidate the physical picture of ultrafast laser-induced structures towards nanoscale based on the viewpoint of the physical regime of plasmonic interaction. Above all, for different regions of the crater irradiated by a Gaussian laser beam, $n_e$ varies in a wide range: at the center $n_e$ can achieve a value much higher than $n_c$, and consequently the over-dense plasma makes the surface strongly metallic; in contrast, at the periphery $n_e$ approaches $n_c$, and thus across the crater boundary a transition from the metallic to the non-metallic behaviour may occur. Accordingly, the physical elements evolve from the center to the periphery: the physical regime converts from the optical regime to the electrostatic regime; the ablation mechanism converts from the non-ultrafast, thermal process to the ultrafast, non-thermal process; the parameters of $n_e$, $\Lambda$, and $W$ decrease and the ratio $D/W$ increases. Correspondingly, the gratings evolve from NSGs to DSGs for the SP-mode conversion, and further scale down approaching nanoscale for stronger electrostatic effects. Moreover, as $N$ increases, the optical regime shrinks along with the NSG declining, and contrarily the electrostatic regime expands along with the DSG growing (see Fig. S5 of Supplemental Material [47]). That is, resulted from the conversion of the physical regime in the multipulse ablation process, the scaling-down of ultrafast laser-induced structures is always a positive feedback process.

## V. CONCLUSIONS

In conclusion, our results demonstrate that in ultrafast laser ablation, the plasmonic effects play an important role for the formation of deep-subwavelength structures. In detail, the SP-mode evolution from the resonant to the nonresonant $K_g$-SP mode and further to the inphase and antiphase $2K_g$-SP modes is responsible for the grating splitting phenomenon. Further, due to the physical regime transforming from the optical regime to the electrostatic regime, the quasistatic SP mode bringing on extraordinary field enhancement should take effect and lead $\Lambda$ and $W$ further scaling down towards the scales of 100 nm and 10 nm, respectively. Besides, the strong electrostatic field enhancement serving as a powerful "evolutionary force" may promote the growth of ultra-regular gratings via multi-pulse feedback. On the whole, in virtue of abundant free electrons produced by ultrafast laser pulses, various plasmonic effects arise spontaneously and make ultrafast laser-induced damage far beyond the diffraction limit—the SP-related mechanisms is of great help in understanding the physical essence of the deep-subwavelength spatial scale and the ultra-short timescale for the formation of ultrafast laser-induced nanostructures: basically, nanocale eliminates electromagnetic retardation effects, facilitates ultrafast dynamics, and arouse tremendous electrostatic field enhancement, and as a result, active plasmonic structures being simultaneously nanoscale and ultrafast arise spontaneously in short-pulse laser ablation.

Moreover, the study suggests an interesting field of natural plasmonics that is not grounded on the elaborate artificial structures produced by human wisdom, but derived from the self-formed structures via the natural-evolution-like process in virtue of the powerful plasmonic "evolutionary force" acting in ultrafast laser ablation. Actually, the various typical plasmonic structures generated in ultrafast laser ablation through the multipulse optimized evolution could present sophisticated, practical examples for certain theoretical predictions in plasmonics, such as the resonant and nonresonant SP modes [31], the quasistatic SP mode [36], and the extraordinary field enhancement effect [36,37], which in turn promote the related investigations through a different perspective.

Furthermore, the ultrafast high-field active plasmonics based on high-excited nanostructured surfaces induced by ultrashort laser pulses should be an emerging field that would give rise to perspectives in other research fields. With regard to the field of active plasmonics [41], one can strongly modulate the dielectric constant ε of a normal material via ultrafast laser irradiation with appropriate conditions. Particularly, semiconductor materials like Si and GaAs, which may gain a great ε change with relatively-low fluence irradiation, are the desired materials for future applications. On the other hand, the extraordinary absorption of high-excited nanostructured surface driven by ultrafast laser would provide inspiration on the field of high-field laser physics [49,50], such as laser fusion and X-ray emission, in which efficient absorption of laser energy for enhancing the generation of fast electrons is extremely important.

## ACKNOWLEDGMENTS

The authors are grateful to Yanfa Liu and Xueran Zeng for their supports in the experiments. This work was supported by National Natural Science Foundation of China (NSFC; grants 11004208 and 11134010), National Basic Research Program of China (grants 2010CB923203 and 2011CB808102), the Open Fund of the State Key Laboratory of High Field Laser Physics (Shanghai Institute of Optics and Fine Mechanics), and the Hundred Talents Program of Sun Yat-Sen University.

## ➢ Supplemental Material for:

# Ultrafast laser-induced subwavelength structures towards nanoscale: the significant role of plasmonic effects

Min Huang,[1,][*] Ya Cheng,[2] Fuli Zhao,[1] and Zhizhan Xu[2]

[1]*State Key Laboratory of Optoelectronic Materials and Technologies, Sun Yat-sen University, Guangzhou 510275, China*
[2]*State Key Laboratory of High Field Laser Physics, Shanghai Institute of Optics and Fine Mechanics, Chinese Academy of Sciences, P.O. Box 800-211, Shanghai 201800, China*
*Corresponding author, email: syshm@163.com

## 1. The example of the grating splitting for a large-area DSG.

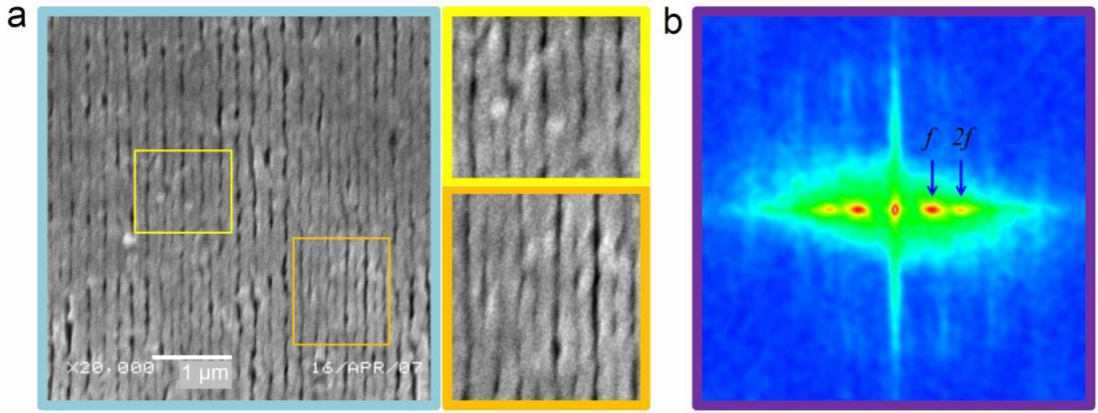

**FIG. S1. The grating-splitting phenomenon for a large-area DSG.** (a) The SEM images of the grating-splitting phenomenon occurring on a large-area DSG produced by 800-nm fs laser with pulse number $N = 3000$ at a fluence of 2.5 J/cm$^2$. Here, the periods ($\Lambda$) of the coarse and fine gratings are about 230 and 115 nm, respectively. (b) The fast fourier transform (FFT) of the square treated area in (a). Mainly, the FFT image reveals

two spatial frequency components: the fundamental frequency (*f* component) corresponds to the original grating with Λ = 230 nm and the second harmonic (2*f* component) corresponds to the splitting grating with Λ = 115 nm (Actually, the non-sinusoidal profile of the original grating provided with a intrinsic second harmonic component would also contribute to the 2*f* component in the FFT image). The result indicates that as long as *N* is large enough, DSGs may further split with the period approaching 100-nm scale.

## 2. The examples of the grating-splitting phenomenon occurring slightly inside the outmost region.

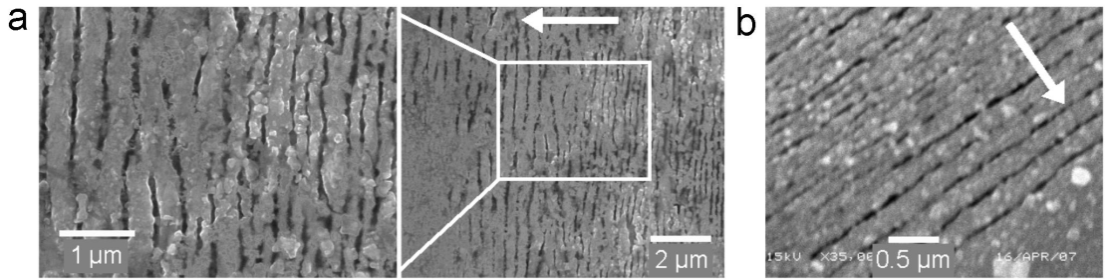

**FIG. S2. The grating-splitting phenomenon occurring slightly inside the outmost region.** (a) SEM images of the peripheral regions of GaAs surface irradiated by 800-nm fs laser with *N* = 1000 at a fluence of 0.17 J/cm². Here, Λ of the coarse and fine gratings are 380 and 190 nm, respectively. (b) SEM images of the peripheral regions of ZnO surface irradiated by 800-nm fs laser with *N* = 5000 at a fluence of 3.0 J/cm². Note that the white arrow points outwards. Here, Λ of the coarse and fine gratings are 220 nm and 105 nm, respectively. These results both show that the gratings splitting tend to occur slightly inside rather than at the outmost region. That is, the coarse gratings always act as the vanguard for the surface structuring process, and adequate feedback is prerequisite for the grating splitting. On the other hand, the results imply that in ultrafast laser ablation the formed gratings are not always in a stable state—as pulse number continues to increase the preformed gratings have the tendency to split and further scale down.

## 3. The evolution of the grating splitting as pulse number increases.

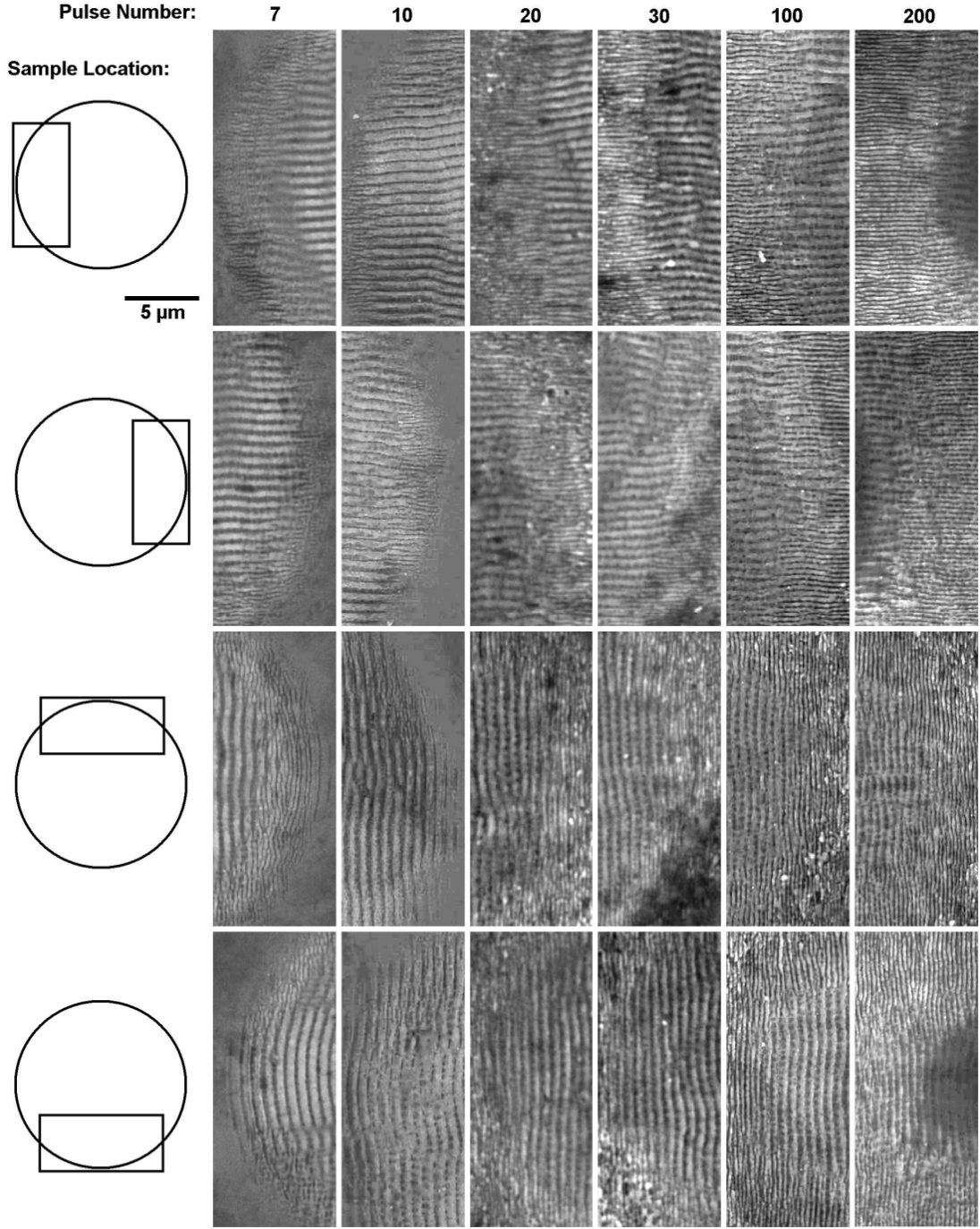

**FIG. S3. The SEM images of the craters produced by a 800-nm fs laser at a fluence of 2.6 J/cm² with different pulse numbers N = 7, 10, 20, 30, 100, and 200.** Here, each crater is divided into four parts toward four directions, thus the corresponding parts can be arranged together to facilitate comparison. All the SEM images have the same scale bar. Obviously, as $N$ increases the conversion from NSGs to DSGs via grating splitting keeps on occurring, along with the conversion regions moving towards the crater center. Besides, in the four directions the splitting has a similar moving trend. Thus, the area of

DSGs rapidly increases, and contrarily the area of NSGs rapidly decreases. Generally, as *N* increases the ultrafast laser-induced structures would spontaneously scale down towards the deep-subwavelength scale.

## 4. The grating-splitting phenomenon for different laser wavelengths.

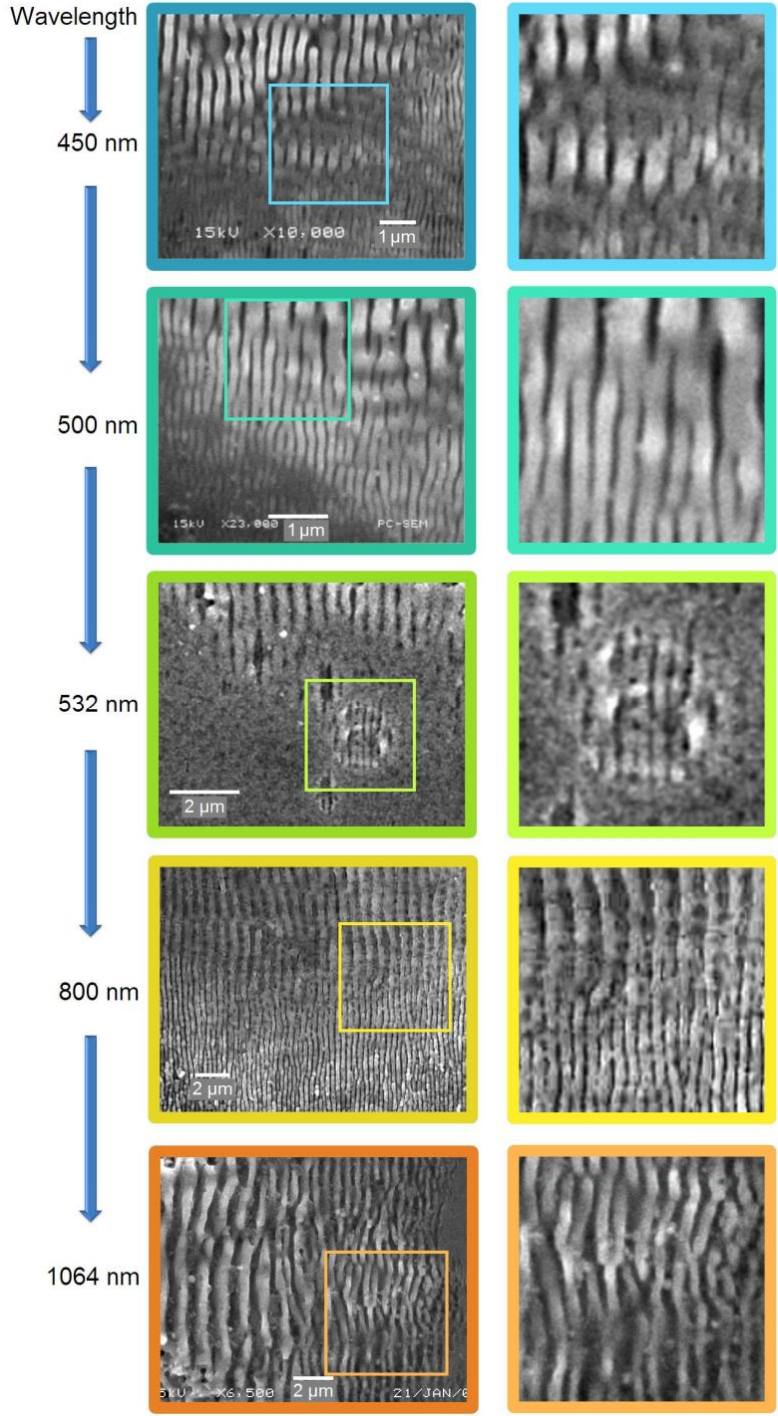

**FIG. S4. The SEM images of the grating-splitting phenomenon on ZnO for different laser wavelength (λ).** The irradiation conditions for the ablated areas from top to bottom are listed as follows (τ-laser duration; F-laser fluence): λ = 450 nm, τ = 30 ps, N = 400, and F = 0.45 J/cm$^2$; λ = 500 nm, τ = 30 ps, N = 300, and F = 0.45 J/cm$^2$; λ = 532 nm, τ = 30 ps, N = 30, and F = 1.63 J/cm$^2$; λ = 800 nm, laser τ = 125 fs, N = 100, and F = 2.6 J/cm$^2$; λ = 1064 nm, τ = 30 ps, N = 200, and F = 5.3 J/cm$^2$. Apparently, the results corresponding to different λ from violet to near-infrared light all exhibit the grating-splitting phenomenon, which indicates the universality of the phenomenon. (Here, the 30-ps pulses of 10-Hz repetition rate were generated by a mode locked Nd:YAG laser system (EKSPLA, PL2143B) equipped with optical parametric generators (OPGs)).

## 5. The evolution of the grating morphologies, physical regimes, and ablation mechanisms as pulse number increases.

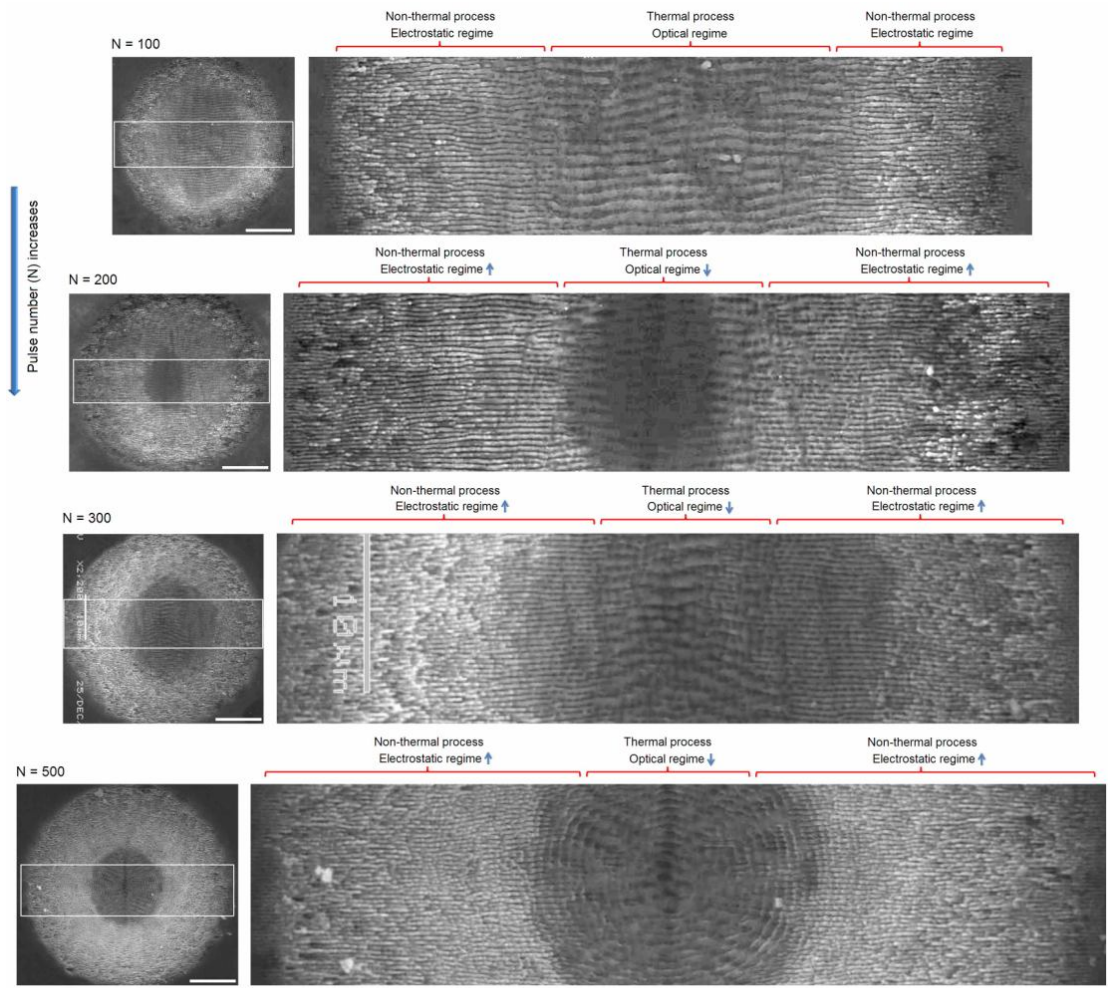

**FIG. S5. The SEM images concerning the evolution of the grating morphologies, physical regimes, and ablation mechanisms for the ablation craters on ZnO as pulse number (*N*) increases.** The craters from top to bottom are produced by 800-nm fs laser at a fluence of 2.6 J/cm$^2$ with pulse number $N$ =100, 200, 300, and 500, respectively. The scale bars all represent 10 μm. By comparing the SEM images corresponding to different pulse number, one can see that as $N$ increases the area of electrostatic regime increases, and contrarily the area of optical regime decreases. Accordingly, the dominant ablation mechanism turns from the non-ultrafast, thermal process to the ultrafast, non-thermal process. Furthermore, these results indicate that as $N$ increases the scaling-down of ultrafast laser-induced structures is always a positive feedback process.